\documentstyle[12pt,a4]{article}

\begin{document}

\title{Trace elements analysis of aerosol samples from some Romanian urban
zones}
\author{I. V. Popescu,
 T. Badica, Agata Olariu\\
{\em National Institute for Physics and Nuclear Engineering}\\
{\em 76900 P. O. Box MG-6, Magurele Bucharest, Romania}}

\maketitle
\begin{abstract}

Aerosols deposits on filters from ten Romanian towns with different kinds and 
levels of industrial development have been studied. The concentration of 
elements with Z$\ge$16 have been
measured by particle-induced X-ray emission (PIXE) analysis. It has been
determined 15 elements: S, K, Ca, Ti, V, Cr, Mn, Fe, Co, Ni, Cu, Zn, As, Hg 
and Pb.

\end{abstract}
\noindent
Keywords: aerosols, pollutants, PIXE\\
\section{Introduction}

In this paper we have analyzed aerosols deposits on filters from ten Romanian 
towns: Pitesti, Giurgiu, Resita, Ramnicu-Valcea, Baia-Mare, Craiova, Timisoara, 
Calarasi, Braila and Arad 
 with different kinds and 
levels of industrial development by method of particle-induced X-ray emission
(PIXE).  This method is based 
on
the fact that the bombardment of the sample charged particles causes
the ionisation of the atomic inner shells followed by a subsequent of the
characteristic X-rays. When the X-rays spectrum is detected by a high resolution
detector, the well-known Z-dependence of the X-rays energies, as well as the
intensities of the individual X-rays line, allow a straight forward 
determination of  elements present in the target.
The properties of PIXE can summarized as: high sensitivity in small samples,
high speed, surface analysis, genuinely multielemental and quantitative,
partialy nondestructive, possible to combine simultaneously with other ion-beam
techniques and microprobes.

The use of protons or alpha particles for the production of inner-shell
vacancies combines a high ionization cross section with low X-ray background.
The background in the region of low-Z elements is determined by the 
bremsstrahlung
from secondary electrons while at higher X-ray energies the background is
normaly determined by $\gamma$-rays produced in the target and the Compton
electrons 
scattered in the crystal of the detector. The selection of various X-rays 
absorbers can
improves the sensitivity over the whole elemental range. Although the
ionization cross sections also increase for high elements with increasing
particle energy up to rather high energies, the variation in the background
radiation leads to the lowest general detection limits being obtained for
1.5-3.5 MeV protons. While the absolute detection limits in thick samples of
low Z-elements are normaly in the interval from 0.1 to 10 $\mu$g/g. The
advantages and disadvantages of the method  as long in use as the PIXE 
analysis are
well known and documented in several reviews, articles and textbooks \cite{1}.

In the present work we report the first PIXE analysis of atmospheric aerosols
deposits filters from ten Romanian towns with different kinds and levels of 
industrialization.

\section{Experimental method}

The experimental set-up was described previously \cite{2}. 
The irradiation chamber
has a 0.2 mm Be window for X-rays. The target was oriented at the 45$^\circ$
angle with respect to both the beam and detector direction. The beam passes
through a collimator ($\Phi$=2mm) before reaching the target. For analysis we
used proton beams of 3 MeV energy supplied by the FN tandem accelerator from
the Institute for Physics and Nuclear Engineering - Bucharest. 
The beam current was
kept below 10 nA to maintain a count rate of about 250 counts/s, which implies
negligible dead-time and pile-up corrections. X-rays were detected with a HPGe
(100 mm$^2\time10$mm) detector with 160 eV energy resolution at 5.9 keV.
Sample targets to be annualized were collected by the Institute of Hydrology and
Waters of Bucharest and prepared in the following manner: 
aerosol particles were
collected on cellulose fiber filter (Whatman 41). The flow rate was  15 to 20
liters per minute. Air volumes were measured with calibrated gasmeters with a
precision of about 5\% (to our regret half of samples have an unknown air
volumes).
The absolute concentrations of observed elements in aerosol samples were
determined by advantage of the internal standard \cite{3}. This calibration
method implies doping the sample with a known amount of the standard element 
and
relating the unknown concentrations to those of the standard element. We choose
Yttrium as the calibration standard because it is very rare element in the 
environment items. The intense peaks of Yttrium in the X-spectrum could obscure
the peaks of some elements possible existing in the samples: L and K lines of
Yttrium 
overlap SK$_{\alpha}$ and (Rb and Sr)K lines respectively. Therefore we have
analyzed targets without Yttrium too and we have not observed any new elements.
A sample of Yttrium on Whatman 41 filter was measured too. Weak impurities of 
Ca, Fe and Zn were found.  Concentrations of elements present in the aerosol
samples were corrected for these impurities of the filter.

\section{Results and discussions}
A typical PIXE spectrum of an air filter sample is shown in Fig. 1. 
\noindent
We have identified 15 elements: S, K, Ca, Ti, V, Cr, Mn, Fe, Co, Ni, Cu, Zn,
As, Hg and Pb.
The
measured elemental concentrations are given in $\mu$g per m$^3$ air in Table 1
for five samples with the known processed air volumes, and with respect to the
concentration of the Ca for all analyzed samples in Table 2. 

The great number of identified elements in the samples is similar to these of
big cities as Livermore (USA) \cite{4} and Munich (Germany) \cite{5}. It is
worth to mention the absence of the Cadmium element in our studied samples.

One could remark from the Table 1 the air filters from the town
Pitesti, for the highest concentrations of the following elements: Ti: 9.63
$\mu$g/m3, V: 1.20 $\mu$g/m3, Ni: 0.59 $\mu$g/m3, S: 7.6 $\mu$g/m3\\
the town
Resita for: Cr: 1.24 $\mu$g/m3, Mn: 10.04 $\mu$g/m3,  Fe:253
$\mu$g/m3,\\
Baia-Mare for Cu: 36 $\mu$g/m3, Zn: 13.5 $\mu$g/m3, As: 5.09
$\mu$g/m3, Pb:12.1 $\mu$g/m3.\\
From the ratios of concentrations shown in the Table 2 we could make a 
comparison
between the analyzed filters from all the  towns considered here, from the
point of view of the pollutant elements: the town Craiova is put in evidence by
its high ratios of concentrations : Ti/Ca: 0.505, Cr/Ca: 0.035, Fe/Ca: 5.38,
Co: 0.015, Zn/Ca: 0.036, As: 0.005 and Pb: 0.043.  Calarasi has the highest 
ratio of Mn/Ca: 0.078  and the filter from Braila is put in evidence by the
presence of Mercury, Hg/Ca:
0.003.
Certainly the level of pollution of a region can not be determined by a single 
filter and it is need of a good statistics to draw conclusions.
This work want rather to demonstrate that the PIXE method is a
suitable tool in the analysis of air filters in the pollution studies and 
and also the results of this analysis draw the attention on the presence of the
pollutants elements from the atmosphere of towns discussed in this paper.

\newpage  
\begin{center}
{\bf Table 1}. Concentrations in air filters, $\mu$g/m$^3$, towns in Romania, by PIXE, with
known processed air volumes \\
\vspace*{1cm}
\begin{tabular}{|c|c|c|c|c|c|}
\hline
\hline
Element&Pitesti&Giurgiu&Resita&Ramnicu-Valcea&Baia-Mare\\
\hline
\hline
S&7.58&-&-&-&-\\
K&38.2&28.7&58.3&-&21.29\\
Ca&67.500&105.6&475&63.9&406.2\\
Ti&9.63&5.33&9.096&3.98&-\\
V&1.200&-&-&0.784&-\\
Cr&0.157&0.733&1.243&0.432&-\\
Mn&1.8&1.49&10.04&11.084&2.44\\
Fe&73.5&51.60&253&41.1&30.41\\
Ni&0.599&-&-&0.410&-\\
Cu&0.238&-&0.344&1.31&36.03\\
Zn&0.802&-&9.14&1.45&13.45\\
As&-&-&-&0.284&5.097\\
Pb&-&-&-&0.369&12.06\\
\hline
\end{tabular}
\end{center} 
\newpage  
\begin{center}
{\bf Table 2}. Ratios of concentrations with respect to
the concentration of the Ca, air filters from industrialzed towns in Romania,
by PIXE\\
\end{center}
\vspace*{1cm}
\scriptsize
\begin{tabular}{|c|c|c|c|c|c|c|c|c|c|c|}
\hline
\hline
Element&Pitesti&Giurgiu&Resita&Ramnicu-Valcea&Baia-Mare&Craiova&Timisoara&Calarasi&Braila&Arad\\
\hline
\hline
S&0.112&-&-&-&-&-&0.02&-&0.012&0.047\\
K&0.57&0.272&0.122&-&0.052&2.51&0.2&0.26&0.174&0.36\\
Ca&1&1&1&1&1&1&1&1&1&1\\
Ti&0.14&0.05&0.019&0.063&-&0.505&0.15&0.66&0.017&0.046\\
V&0.018&-&-&0.012&-&-&-&-&-&-\\
Cr&0.002&0.007&0.003&0.007&-&0.035&-&0.12&0.006&0.001\\
Mn&0.027&0.014&0.021&0.017&0.006&0.05&0.007&0.078&0.024&0.017\\
Fe&1.09&0.49&0.532&0.647&0.075&5.38&0.52&4.32&0.28&0.58\\
Co&-&-&-&-&-&0.015&-&-&-&-\\
Ni&0.009&-&-&0.006&-&-&-&-&-&-\\
Cu&0.004&-&0.001&0.021&0.089&-&0.001&0.008&0.002&0.03\\
Zn&0.012&-&0.019&0.023&0.033&0.036&0.002&0.015&0.004&0.011\\
As&-&-&-&0.004&0.004&0.005&-&-&-&0.001\\
Hg&-&-&-&-&-&-&0.001&-&0.003&0.001\\
Pb&-&-&-&0.006&0.03&0.043&-&0.005&0.002&0.0003\\
\hline
\end{tabular}
  
\end{document}